\documentclass[letterpaper]{article} 
\usepackage{aaai2026}  
\usepackage{times}  
\usepackage{helvet}  
\usepackage{courier}  
\usepackage[hyphens]{url}  
\usepackage{graphicx} 
\usepackage{natbib}  
\usepackage{caption} 
\usepackage{algorithm}
\usepackage{algorithmic}
\usepackage{amsmath} 
\usepackage{amsmath}
\usepackage{newfloat}
\usepackage{listings}
\usepackage{booktabs} 
\DeclareCaptionStyle{ruled}{labelfont=normalfont,labelsep=colon,strut=off} 
\floatstyle{ruled}
\newfloat{listing}{tb}{lst}{}
\floatname{listing}{Listing}
\title{MF-Speech: Achieving Fine-Grained and Compositional Control in Speech Generation via Factor Disentanglement}
\author{
    Xinyue Yu\textsuperscript{\rm 1},
    Youqing Fang\textsuperscript{\rm 1},
    Pingyu Wu\textsuperscript{\rm 1}\equalcontrib,
    Guoyang Ye\textsuperscript{\rm 1}, \\
    Wenbo Zhou\textsuperscript{\rm 1}, 
    Weiming Zhang\textsuperscript{\rm 1}\equalcontrib,
    Song Xiao\textsuperscript{\rm 2}
}
\affiliations{
    \textsuperscript{\rm 1}School of Information Science and Technology, University of Science and Technology of China, Hefei 230026, China.\\
    \textsuperscript{\rm 2}Department of Electronic and Communication
 Engineering, Beijing Electronic Science and Technology Institute, Beijing
 100070, China.\\

    \{xinyueyu, fangyq, wupingyu, studywithme\}@mail.ustc.edu.cn,\\
    \{welbeckz, zhangwm\}@ustc.edu.cn, xs\_xidian@163.com
}

\usepackage{bibentry}

\begin{document}

\maketitle

\begin{abstract}
Generating expressive and controllable human speech is one of the core goals of generative artificial intelligence, but its progress has long been constrained by two fundamental challenges: the deep entanglement of speech factors and the coarse granularity of existing control mechanisms. To overcome these challenges, we have proposed a novel framework called MF-Speech, which consists of two core components: MF-SpeechEncoder and MF-SpeechGenerator. MF-SpeechEncoder acts as a factor purifier, adopting a multi-objective optimization strategy to decompose the original speech signal into highly pure and independent representations of content, timbre, and emotion. Subsequently, MF-SpeechGenerator functions as a conductor, achieving precise, composable and fine-grained control over these factors through dynamic fusion and Hierarchical Style Adaptive Normalization (HSAN). Experiments demonstrate that in the highly challenging multi-factor compositional speech generation task, MF-Speech significantly outperforms current state-of-the-art methods, achieving a lower word error rate (WER=4.67\%), superior style control (SECS=0.5685, Corr=0.68), and the highest subjective evaluation scores (nMOS=3.96, sMOS$_t$=3.86, sMOS$_e$=3.78). Furthermore, the learned discrete factors exhibit strong transferability, demonstrating their significant potential as a general-purpose speech representation.
\end{abstract}

\begin{links}
    \link{Demo}{https://guoyang25.github.io/mf-speech/}
\end{links}

\section{Introduction}
Infusing life into the digital world and endowing speech with personality and emotion is one of the most exciting frontiers in the field of generative artificial intelligence. From emotionally aware assistants to personalized voice restoration and expressive media synthesis \cite{sisman2020overview,zhang2019joint,veaux2013towards}, it is poised to transform how we interact with the digital world. Voice Conversion \cite{bargum2023reimagining} enables flexible manipulation of fundamental speech factors such as content, timbre, and emotion. As such, Voice Conversion has emerged as a key enabling technology toward this vision. However, two fundamental challenges have long troubled researchers in this field:

\begin{figure}[t]
  \centering
  \includegraphics[width=\columnwidth]{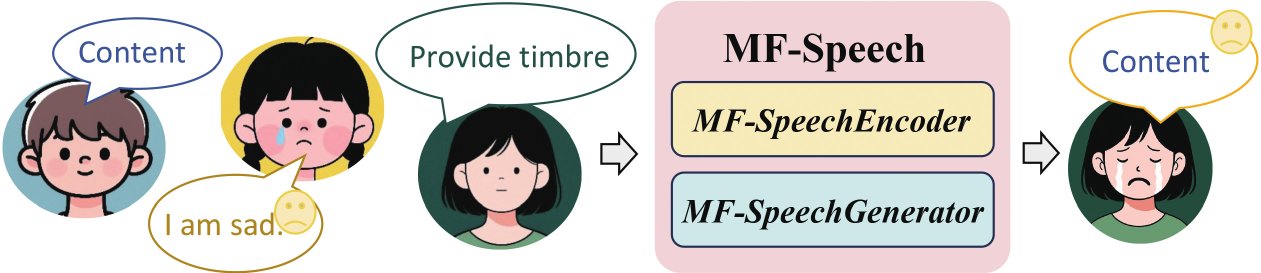}
  \caption{MF-Speech enables independent and fine-grained control over speech content, timbre, and emotion factors for speech synthesis.}
  \label{fig:1_Teaser}
\end{figure}

\begin{itemize}
\item  \textbf{Gene hybridization: The Challenge of Pure Factor Separation.} The content, timbre and emotion in speech are naturally intertwined and hard to separate. Due to the lack of a strong supervisory signal, the current strategies \cite{chou2018multi, yadav2023dsvae, qian2020unsupervised, li2023styletts, wang2021adversarially, wang2018style, li2025styletts, chou2019one} act like rough filters, making it difficult to precisely and accurately separate various speech factors and thereby leading to timbre leakage and attribute interference. Moreover, this deep entanglement can also lead to fragile and chaotic factor representations, not only disrupting the precise control of each factor but also severely limiting their transferability across tasks \cite{li2023dvqvc, yuan2021improving, lian2022robust, mu2024self, tu2024contrastive, song2022multi, deng2024learning}. 

\item  \textbf{Command failure: The Lack of Fine-Grained Control.} Even if pure speech factors are obtained, how to skillfully control them remains a major challenge. The existing control mechanisms are generally coarse-grained, just like using a sledgehammer to complete fine carving work. Whether models rely on the fundamental methods like static concatenation and implicit global modulation \cite{kaneko2019stargan, qian2019autovc, neekhara2023selfvc, zhang2019non, yao2024promptvc}, or employ advanced technologies such as dynamic fusion and explicit modulation \cite{yao2025stablevc,ma2024vec,ning2023expressive, choi2025voiceprompter,li2024sef,qian2020f0}, they consistently suffer from coarse-grained control. This is due to their failure to systematically combine dynamic weights and hierarchical style injection. Therefore, models often struggle to balance content fidelity (content) and style similarity (timbre and emotion). This fundamental defect cannot be remedied through post-processing techniques \cite{ren2020fastspeech, lin2024voxgenesis, tian2024user}.
\end{itemize}

To address these two major challenges, we propose MF-Speech, a framework designed to achieve fine-grained and compositional control in speech generation via multi-factor disentanglement (Figure~\ref{fig:1_Teaser}). This framework consists of \textbf{Multi-factor Speech Encoder (MF-SpeechEncoder)} and \textbf{Multi-factor Speech Generator (MF-SpeechGenerator)}. It fundamentally addresses the aforementioned challenges by enhancing the purification capability and clarifying the command direction, achieving composable and fine-grained control speech generation. Our main contributions can be summarized as follows:

\begin{itemize}
    \item \textbf{Multi-factor Speech Encoder to ensure factor purity (MF-SpeechEncoder).} We designed a speech factor purifier that uses a three-stream architecture and decomposes the raw speech signal into three highly pure and mutually independent information streams: content, timbre, and emotion. This ensures a high degree of independence for subsequent control and addresses the challenge of gene hybridization.
    \item \textbf{Multi-factor Speech Generator to enhance control granularity (MF-SpeechGenerator).} Building upon the purified factors, we developed the speech factor conductor. This component moves beyond coarse control by incorporating dynamic fusion and Hierarchical Style Adaptive Normalization (HSAN). This enables highly fine-grained control over timbre and emotion. As a result, the model can synthesize a vast array of speech combinations with high style similarity, while maintaining content fidelity.
    \item \textbf{Comprehensive empirical and systematic validation.} Extensive experiments demonstrate the effectiveness of our proposed framework. Results show that MF-SpeechEncoder can effectively purify speech factors to ensure control independence. Moreover, in the challenging task of multi-factor compositional speech generation, the MF-Speech demonstrates remarkable controllability in terms of content fidelity and style similarity.
\end{itemize}

\section{Related Work}
\textbf{Factor Disentanglement: Strategies and Challenges.} VQMIVC \cite{wang2021vqmivc} separates pitch, content, and timbre through vector quantization and mutual information minimization. However, F0 is not explicitly modeled but rather provided as an external condition, leading to unnatural prosody in converted speech. Moreover, as pitch is a fundamental acoustic feature, it is inherently entangled with timbre, resulting in implicit pitch–timbre entanglement and residual timbre leakage. StyleVC \cite{hwang2022stylevc} uses a global style encoder that outputs a single vector, making it difficult to disentangle timbre and emotion. Prosodic information is only supplemented through external auxiliary features, and adversarial training is limited to content–style separation, allowing emotion and timbre to remain entangled within the style representation. StableVC \cite{yao2025stablevc} incorporates a gradient reversal layer (GRL) on top of FaCodec’s style component to further disentangle style and timbre factors. However, since FaCodec is primarily optimized for F0 modeling, emotion factor remains implicitly entangled and is not explicitly modeled. In summary, existing methods often suffer from impure factor definitions, limited architectural design, and limited training objectives, resulting in incomplete disentanglement. In contrast, we explicitly model emotion factor based on prosody information and design dedicated modules for timbre, emotion, and content under a unified multi-objective optimization framework. This enables reduced mutual interference among the factors and enhances the ability to control them independently.

\textbf{Conditional Generation: Mechanisms and Granularity.} MSM-VC \cite{wang2023msm} and DDDM-VC \cite{choi2024dddm} employ static feature concatenation and inject it into the decoder, which offers limited flexibility in style control. NS2VC \cite{shen2023naturalspeech} uses a FiLM-based affine transformation driven by speech prompts, achieving a degree of dynamic control, yet its mechanism remains less flexible than a fully disentangled, hierarchical design. Facodec \cite{ju2024naturalspeech} simulates dynamic control via serial prompt concatenation, but this essentially functions as an advanced form of static fusion, lacking real-time adjustment of factor weights. Although StableVC \cite{yao2025stablevc} leverages conditional flow matching and attention mechanisms to improve dynamic feature fusion—thereby enhancing the independent control of timbre and style—it still relies on relatively static generation control with a shallow overall structure and lacks explicit dynamic weighting of factor contributions. HierVST \cite{lee2023hiervst}  presents a hierarchical style injection mechanism, a structural advancement, but it lacks dynamic factor weighting and explicit balancing of factor contributions. In summary, existing control mechanisms are typically designed in a static, global, or non-collaborative manner, which limits their capacity for fine-grained and compositional control. In contrast, our framework employs a dynamic fusion module and HSAN. This approach not only considers the adaptive weighting of each factor but also enables style injection in each layer in a fine-grained manner, significantly enhancing the flexibility of multi-factor composition.

\section{MF-Speech}
\label{sec:method}
\subsection {Overview}

\begin{figure*}[t]
  \centering
  \includegraphics[width=0.9\textwidth]{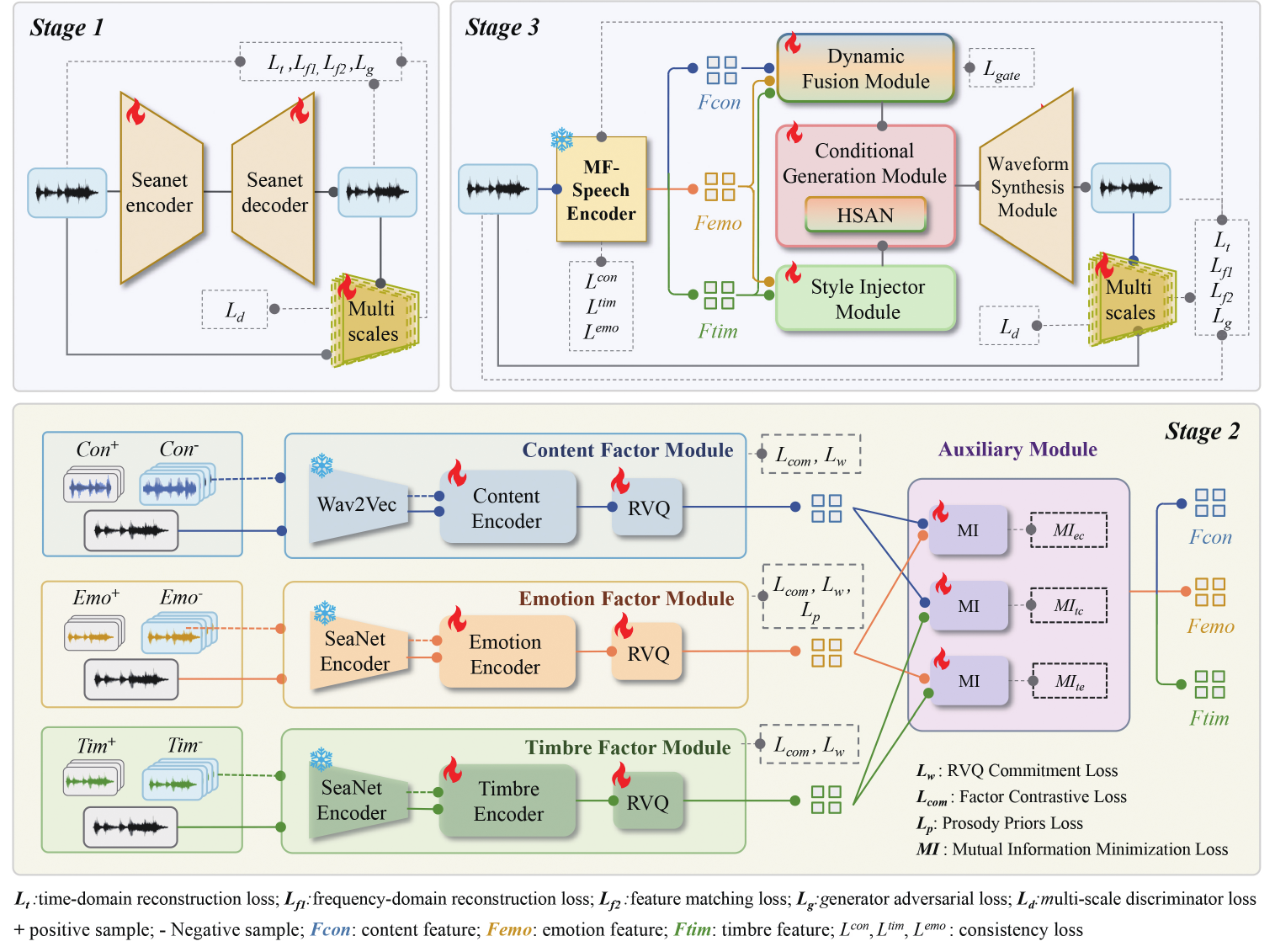}
  \caption{The training process of MF-Speech consists of three stages. The first stage ensures high-precision conversion between waveforms and features. The second stage disentangles clean and independent content, timbre, and emotion factors. The third stage enables fine-grained, multi-factor control for waveform generation.}
  \label{fig:A1_trainProcess}
\end{figure*}

To enable more independent, fine-grained, and controllable speech synthesis, we propose MF-Speech (Figure~\ref{fig:A1_trainProcess}). It extracts three independent and high-purity speech factors. Then, it dynamically integrates these factors and injects the timbre and emotion at each layer. Thereby, it improves the content fidelity and style similarity of the generated speech. MF-Speech consists of two core components:

\begin{itemize}
    \item  \textbf{MF-SpeechEncoder}: A high-purity factor encoder, which extracts disentangled discrete representations for each core factor from input speech.
    \item  \textbf{MF-SpeechGenerator}: A fine-grained waveform generator, which synthesizes the final speech by precisely controlling the fusion and modulation based on arbitrary combinations of discrete factor representations.
\end{itemize}

\subsection{MF-SpeechEncoder}
The design of MF-SpeechEncoder aims to learn high-purity and mutually independent representations of speech content, timbre, and emotion, which is an essential foundation for fine-grained and compositional control in speech
generation. As illustrated in Figure~\ref{fig:A1_trainProcess}, MF-SpeechEncoder is trained in stage 2 under a multi-objective optimization strategy. The framework adopts a three-stream architecture, consisting of three specialized submodules and an information-theoretic auxiliary module.

\begin{itemize}
    \item \textbf{Content Factor Module:} To isolate pure speech content, this module first extracts initial representations using a pre-trained Wav2Vec2 \footnote{\url{https://huggingface.co/facebook/wav2vec2-base-960h}} \cite{baevski2020wav2vec} model as its backbone. A lightweight trainable sub-network then refines these representations via sentence-level content contrastive learning, specifically to suppress residual timbre and emotion information. Finally, a Residual Vector Quantizer (RVQ) \cite{yang2023hifi} discretizes these purified representations into discrete content tokens.
    
    \item \textbf{Emotion Factor Module:} Recognizing that emotional expression heavily relies on prosody dynamics, this module adopts a two-stage architecture. Initially, lightweight predictors explicitly generate F0 and energy representations from an intermediate layer, guided by direct supervision to focus on emotion-related acoustic cues. The final emotion representation is then derived from these predicted prosody representations. An emotion contrastive loss enhances the discriminability of these representations across different emotional states, which are subsequently discretized using RVQ to yield controllable and transferable emotion representations.

    \item \textbf{Timbre Factor Module:} This module aims to create stable and generalizable timbre representations. After a SeaNet encoder \cite{tagliasacchi2020seanet}, it employs a multi-head attention mechanism \cite{deora2023optimization} to aggregate and enhance global timbre representations from the input speech. To further purify these representations from content and emotion interference and bolster representation robustness, a timbre-specific contrastive loss is applied. The resulting representations are then discretized using RVQ.

    \item \textbf{Information Theory Constraints:} To mitigate potential residual entanglement between the outputs of the dedicated factor modules, we apply structural regularization using mutual information (MI) minimization after discretization. Inspired by MAIN-VC \cite{li2024main}, a separate MI estimation network employing CLUB \cite{cheng2020club} and MINE \cite{belghazi2018mine} are trained to penalize redundant information between the factor representations, thereby promoting their independence. This MI estimation network is active only during training to maintain operational efficiency at inference.

    \item \textbf{Optimization Objectives:} The MF-SpeechEncoder is trained to produce disentangled discrete representations of content, timbre, and emotion using a multi-objective loss function. This total loss, $\mathcal{L}_{\text{Encoder}}$, combines objectives for RVQ $\mathcal{L}_{w}^f$, factor-specific contrastive learning $\mathcal{L}_{\text{com}}$, prosody priors $\mathcal{L}_{\text{p}}$, and mutual information (MI) minimization constraints $\mathcal{L}_{\text{MI}}$. The MI constraints are introduced gradually via a warm-up schedule to avoid hindering initial representation learning. The overall MF-SpeechEncoder objective is shown in Equation ~\ref{eq:loss_de}, where $t,e,c$ represent content, timbre, and emotion factors, respectively, and $\alpha$(epoch) is the warm-up weighting for MI loss. See the \textit{Appendix A} for more details.
\begin{equation} 
\begin{split}
\mathcal{L}_{\text{Encoder}} =\ &
\sum_{f \in \{\text{t}, \text{e}, \text{c}\}} \lambda_{com}^f \cdot \mathcal{L}_{\text{com}}^f +\sum_{f \in \{\text{t}, \text{e}, \text{c}\}} \lambda_{w}^f \cdot \mathcal{L}_{w}^f \\ 
& + \lambda_p \cdot \mathcal{L}_{\text{p}} + \alpha(\text{epoch}) \cdot \sum_{\text{X,Y}} \mathcal{L}_{\text{MI}}(\text{X,Y}).
\end{split}
\label{eq:loss_de} 
\end{equation}
\end{itemize}

\subsection {MF-SpeechGenerator}
\begin{figure}[t]
  \centering
  \includegraphics[width=\columnwidth]{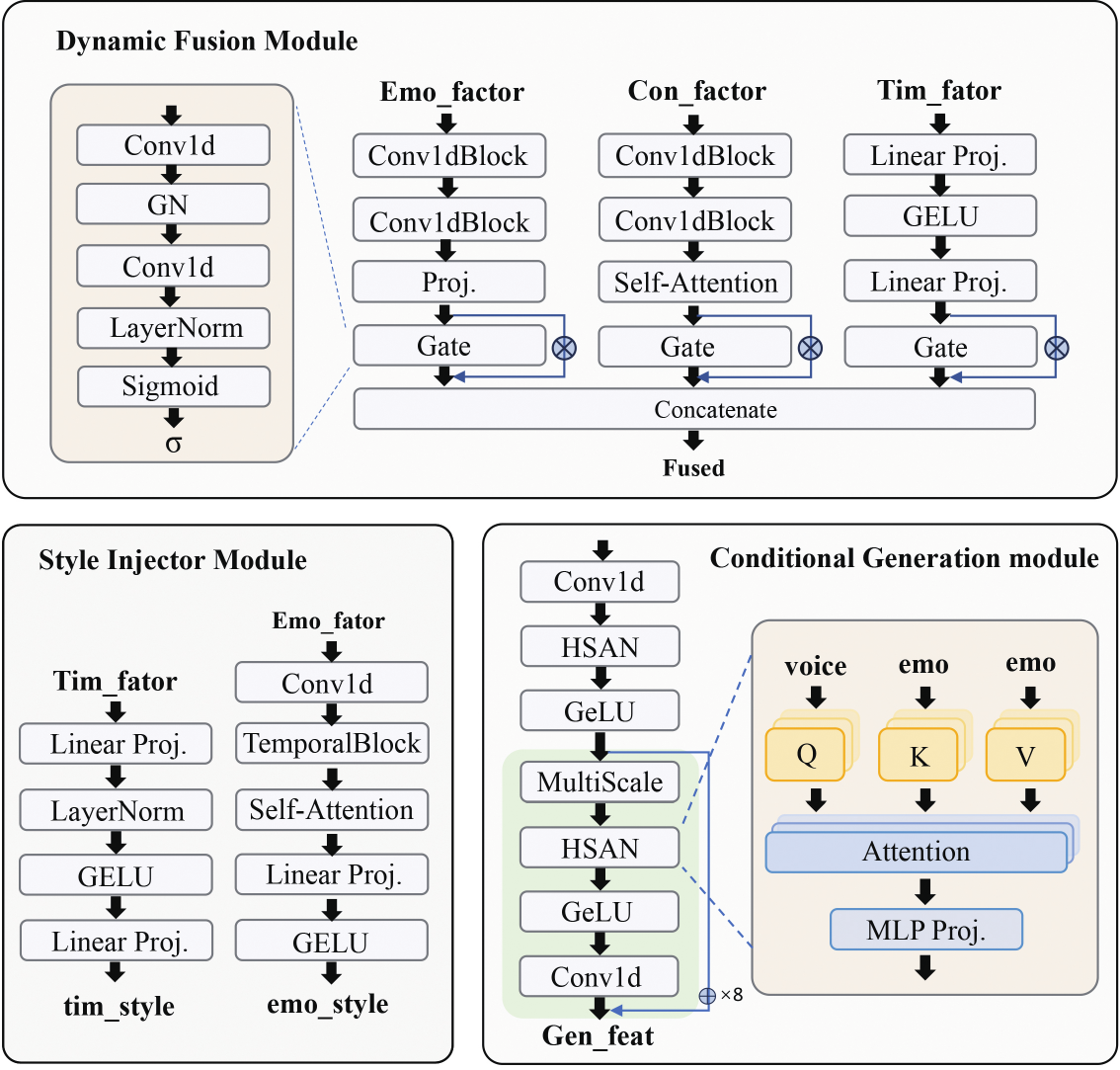}
  \caption{The related architectures and data flows of dynamic fusion and HSAN}
  \label{fig:A3_CSG}
\end{figure}

MF-SpeechGenerator performs fine-grained and compositional control in Speech
Generation based on the discrete factor representations provided by MF-SpeechEncoder, through dynamic fusion and HSAN. As shown in Figure~\ref{fig:A1_trainProcess}, this process is trained in stage 3, which comprises four collaboratively functioning modules: dynamic fusion, style injection, conditional generation, and waveform synthesis. Dynamic fusion and HSAN are shown in Figure~\ref{fig:A3_CSG}.

\begin{itemize}
    \item \textbf{Dynamic Fusion Module:} This initial module dynamically integrates the discrete content, timbre, and emotion representations from the MF-SpeechEncoder into a unified conditional representation for subsequent acoustic modeling. It employs a dynamic gating mechanism that generates time-varying weights for each factor. This adaptive, weighted integration allows the model to flexibly regulate the influence of each factor at different time steps during synthesis.

    \item \textbf{Style Injection Module:} To enable fine-grained control over timbre and emotion, this module is essentially a style parameter generator. It infers multi-level style parameter representations from the discrete representations of timbre and emotion. This facilitates their integration into the generator to implement the HSAN mechanism.
    
    \item \textbf{Conditional Generation Module:} This module constructs highly conditional temporal acoustic representations. It maps the dynamically fused factor representations into detailed acoustic representation sequences, guided by the hierarchical style parameters from the style parameter generator. The backbone network uses stacked residual blocks integrated with multi-scale convolution modules to effectively capture complex temporal dependencies. Crucially, the HSAN layer is applied across multiple levels of this network. At each level, style parameters adaptively modulate the representations, ensuring the acoustic generation process is consistently conditioned and precisely controlled by fine-grained style information. HSAN first fuses timbre and emotion representations via cross-attention. This fused result is projected to yield affine parameters ($\gamma$, $\beta$) and a residual modulation term ($\alpha$). Equation~\ref{eq:san} defines a transformation that combines an affine function with residual modulation, where $IN(x)$ denotes instance normalization, $x$ is the input feature, and $\lambda$ is a scalar. This formulation allows normalized features to be scaled and shifted, with an added expressive residual term enhancing stylistic control.
    \begin{equation}
    \label{eq:san}
    \mathbf{y} = \mathrm{IN}(\mathbf{x})(1 + \tanh(\boldsymbol{\gamma})) + \boldsymbol{\beta} + \lambda \tanh(\alpha) \odot \mathbf{x}.
    \end{equation}

    \item \textbf{Waveform Synthesis Module:} Finally, this module converts the generated acoustic representation sequence into the speech waveform. We employ the SeaNet decoder \cite{tagliasacchi2020seanet} as the waveform decoder. Its robust capability ensures preservation of acoustic details and high perceptual quality. During training, it is initially frozen and then fine-tuned via a phased unfreezing strategy, allowing it to adapt optimally to the generator's output and further enhance synthesis quality.
    
     \item \textbf{Optimization Objectives:} The MF-SpeechGenerator is trained using adversarial learning, with a phased unfreezing strategy for the pre-trained waveform decoder, to enhance realism and ensure stable convergence. The generator's composite loss function ($\mathcal{L}_{\text{Generator}}$) is shown in Equation ~\ref{eq:loss_csg}. See the \textit{Appendix B} for more details.
\begin{equation} 
\begin{split}
\mathcal{L}_{\text{Generator}} = \ & \lambda_{\text{gate}} \mathcal{L}_{\text{gate}} + \lambda_{g} \mathcal{L}_{g}  + \lambda_{\text{feat}} \mathcal{L}_{\text{feat}} \\
& + \lambda_{t} \mathcal{L}_{t} + \lambda_{f} \mathcal{L}_{f} + \lambda_{\text{sim}} \mathcal{L}_{\text{sim}}.
\end{split}
\label{eq:loss_csg}
\end{equation}
The multi-scale discriminator \cite{defossez2022high} loss ($\mathcal{L}_d$) is shown in Equation ~\ref{eq:loss_disc}. It is a hinge loss to distinguish real speech $x$ from generated speech $\hat{x}$. 
\begin{equation}
\label{eq:loss_disc}
\mathcal{L}_d = \frac{1}{K} \sum_{k=1}^{K} \left[ \max(0, 1 - D_k(x)) + \max(0, 1 + D_k(\hat{x})) \right].
\end{equation} 
\end{itemize}

\section{Experiments}
\subsection {Experimental Setup}

\textbf{Dataset and Training Details:} Since the ESD dataset \cite{zhou2021seen} contains explicit emotion labels, we selected it for our experiments and divided it into a training set, a seen test set, and an unseen test set. The MF-Speech framework was trained in three stages on a single NVIDIA 4090 GPU. Stage 1 was trained for 92,000 iterations with a batch size of 24, $\lambda_{g}$=3, $\lambda_{feat}$=3, $\lambda_{t}$=0.1, $\lambda_{f}$=1. Stage 2 was trained for 27,500 iterations with a batch size of 12, $\lambda_{com}$=5, $\lambda_{w}$=1, $\lambda_{p}$=2. Stage 3 was trained for 91,800 iterations with a batch size of 72, $\lambda_{gate}$=1, $\lambda_{sim}$=1, $\lambda_{g}$=3 , $\lambda_{feat}$=3, $\lambda_{t}$=0.1, $\lambda_{f}$=1.

\textbf{Baseline Systems:} To benchmark the effectiveness of MF-Speech in fine-grained and compositional control in speech generation, we select four representative baselines that cover the main paradigms in this field. \textbf{StyleVC} \cite{hwang2022stylevc} leverages adversarial training to achieve disentanglement between content and style, demonstrating strong adaptability to non-parallel data. \textbf{NS2VC} \cite{shen2023naturalspeech} introduces a latent diffusion mechanism to implicitly model global style representations. \textbf{FaCodec} \cite{ju2024naturalspeech} employs explicit factor decomposition, offering high structural interpretability and fine-grained control at the factor level. \textbf{DDDM-VC}  \cite{choi2024dddm} adopts an iterative diffusion process with multi-modal conditioning to enable fine-grained and flexible style control. By comparing with these baselines, we can verify MF-Speech's independence and fine-grained control capabilities.

\subsection {Evaluation Overview}
\textbf{Evaluation Scheme:} The input data is constructed as content ($c$), timbre ($t$), and emotion ($e$). When $ c = e = t $, it corresponds to the speech reconstruction task, generating reconstruction data. If any one of them differs, it corresponds to the multi-factor compositional speech generation task, producing controllable data. Using this setup, we generated 200 reconstruction samples and 200 controllable samples with both the MF-Speech and baseline methods, and evaluated the task performance accordingly. To evaluate the independence of each factor, we assess performance on target and non-target tasks, compute mutual information between factor pairs, and use t-SNE to visualize the factor representations. The specific calculation details of the indicators are in \textit{Appendix C}.

\textbf{Subjective Evaluation:} We selected 20 participants for subjective evaluation. \textbf{nMOS:} Overall perceived naturalness of the generated speech. \textbf{sMOS$_t$, sMOS$_e$:} Similarity of timbre (sMOS$_t$) and emotion (sMOS$_e$) to reference speech.

\textbf{Objective Evaluation.} \textbf{MI:} Quantify redundancy between the learned discrete representations of content ($c$), timbre ($t$), and emotion ($e$). \textbf{Acc:} Accuracy on target tasks (e.g., speaker ID from timbre) and non-target tasks (e.g., emotion ID from timbre) to measure information leakage. \textbf{t-SNE:} Used to qualitatively illustrate the clustering and separability of the learned factor representations. \textbf{UTMOS \cite{saeki2022utmos}:} An objective MOS prediction system\footnote{\url{https://github.com/tarepan/SpeechMOS}}. \textbf{SECS:} Cosine similarity between speaker embeddings of generated and reference samples. \textbf{Log RMSE, Corr:} Log Root Mean Squared Error (Log RMSE) and the Pearson Correlation Coefficient (Corr) for F0 contours, aligned using Dynamic Time Warping (DTW) \cite{muller2007dtw}. These are used to evaluate style similarity, as emotion is highly correlated with F0. \textbf{WER:} Word Error Rate, calculated by comparing ground truth transcriptions with those from a pre-trained Automatic Speech Recognition (ASR) model \footnote{\url{https://github.com/speechbrain/speechbrain}} applied to the generated speech.

\begin{table*}[h]
  \centering
  \begin{tabular}{lcccccccc}
    \hline
    \multicolumn{9}{c}{\textbf{Speech Reconstruction}} \\
    \hline
    Name & nMOS$\uparrow$ & sMOS$_t\uparrow$ & sMOS$_e\uparrow$ & $UTMOS\uparrow$ & SECS $\uparrow$ & Log RMSE$\downarrow$ & Corr$\uparrow$ & WER$\downarrow$  \\ 
    \hline
    GT & -  & - & - & 3.9193 & - & -  & - & -     \\
    StyleVC     & 3.54 $\pm$ 0.15  & 3.56  $\pm$  0.14 & 3.32  $\pm$  0.16 &  3.4536 & 0.2015 & 0.23 &0.5 & 28.16\%    \\
    NS2VC    & 2.67 $\pm$ 0.17  & 2.52  $\pm$  0.18  & 2.63  $\pm$  0.17 &  2.3982 & 0.2758 & 0.12 & 0.87 & 7.17\%  \\
    DDDM-VC     & \textbf{3.87 $\pm$ 0.11}  & \textbf{3.86 $\pm$ 0.12} & \textbf{3.79 $\pm$ 0.10} & \textbf{3.6116} & 0.6430 & \textbf{0.03} & \textbf{0.99} & 5.17\%    \\
    FACodec     & 2.30 $\pm$ 0.15  & 2.16  $\pm$  0.16 & 2.27  $\pm$  0.15 & 1.9146 & 0.3002 & 0.09 & 0.92 & 18.17\%   \\
    \textbf{MF-Speech}     & 3.53 $\pm$ 0.13  & 3.54  $\pm$  0.13 & 3.50  $\pm$  0.13 & 2.7314 & \textbf{0.7401} & 0.08 & 0.94 & \textbf{2.83\%}   \\
    \hline
    \multicolumn{9}{c}{\textbf{Multi-factor Compositional Speech Generation}}\\
    \hline
     Name & nMOS$\uparrow$ & sMOS$_t\uparrow$ & sMOS$_e\uparrow$ & UTMOS$\uparrow$ & SECS $\uparrow$ & Log RMSE$\downarrow$ & Corr$\uparrow$ & WER$\downarrow$  \\ 
    \hline
    StyleVC   &  2.81 ± 0.31 & 2.98 ± 0.33 & 2.40 ± 0.36 & \textbf{3.2176}  & 0.0985 & 0.35 & 0.48 & 24.83\%    \\
    NS2VC    & 3.76 ± 0.26 & 3.11 ± 0.31 & 3.44 ± 0.25 & 2.0270  & 0.1552 & 0.43 & 0.55 & 23.33\%   \\
    DDDM-VC   & 3.58 ± 0.34
& 3.50 ± 0.26
& 3.13 ± 0.33
 & 2.8388  &  0.3723 & 0.37 & 0.62 & 11.67\%    \\
    FACodec   & 2.83 ± 0.34
& 2.38 ± 0.34
& 3.14 ± 0.30
& 1.7128  & 0.1866 & 0.41 & 0.58 & 29.17\%    \\
    \textbf{MF-Speech}  &  \textbf{3.96 ± 0.31}
&  \textbf{3.86 ± 0.30}
&   \textbf{3.78 ± 0.27}
 & 2.6686  & \textbf{0.5685} & \textbf{0.34} & \textbf{0.68} & \textbf{4.67\%} \\ 
    \hline
  \end{tabular}
\caption{The subjective and objective evaluation results of MF-Speech and baseline systems and ``GT" refers to the real sample.}
\label{tab:reconstruction}
\end{table*}

\subsection {Experimental Results on MF-Speech}
We evaluated MF-Speech on the speech reconstruction task and the multi-factor compositional speech generation task, with detailed results presented in Table \ref{tab:reconstruction}. Overall, MF-Speech demonstrates strong control, as measured by metrics for content fidelity and style similarity.

\textbf{Speech Reconstruction:} MF-Speech achieves the highest timbre similarity (SECS = 0.7401) and the lowest word error rate (WER = 2.83\%), demonstrating strong control over both timbre and content. Its emotion consistency is also competitive, with a high F0 correlation (Corr = 0.94) and a low F0 reconstruction error (Log RMSE = 0.08), closely approaching the performance of the best-performing model, DDDM-VC. In terms of subjective naturalness (nMOS = 3.53) and style similarity (sMOS$_t$ = 3.54; sMOS$_e$ = 3.50), MF-Speech performs comparably to StyleVC and is only slightly behind DDDM-VC. The perceptual quality of the generated speech may have influenced similarity judgments to some extent. Overall, while DDDM-VC slightly outperforms MF-Speech across several metrics, the differences are relatively minor. Notably, the factor controllability and disentanglement demonstrated in the reconstruction task remain limited.

\textbf{Multi-factor Compositional Speech Generation:} To further evaluate controllability, we conducted a more challenging experiment on multi-factor compositional speech generation, where content, timbre, and emotion are combined from different sources. Apart from UTMOS, MF-Speech achieved the best performance across all other metrics (SECS = 0.5685, Log RMSE = 0.34, Corr = 0.68, and WER = 4.67\%). Although MF-Speech achieves a slightly lower UTMOS score than StyleVC, the noticeably faster speaking rate of StyleVC negatively affects its perceived naturalness, resulting in a higher subjective evaluation score for MF-Speech. These results demonstrate that MF-Speech can successfully generate speech with accurate content and high similarity in both timbre and emotion, highlighting its strong capability in controllable and compositional generation.

\textbf{Summary of MF-Speech Performance:}
These results demonstrate that MF-Speech achieves fine-grained and compositional speech generation. The generated speech has high fidelity (content) and style similarity (timbre and emotion). Although slightly behind DDDM-VC in reconstruction, MF-Speech excels in multi-factor compositional generation, showing strong control over content, timbre, and emotion.

\subsection {Experimental Results on MF-SpeechEncoder}
\begin{figure*}[htbp]
  \centering
    \includegraphics[width=1.0\textwidth]{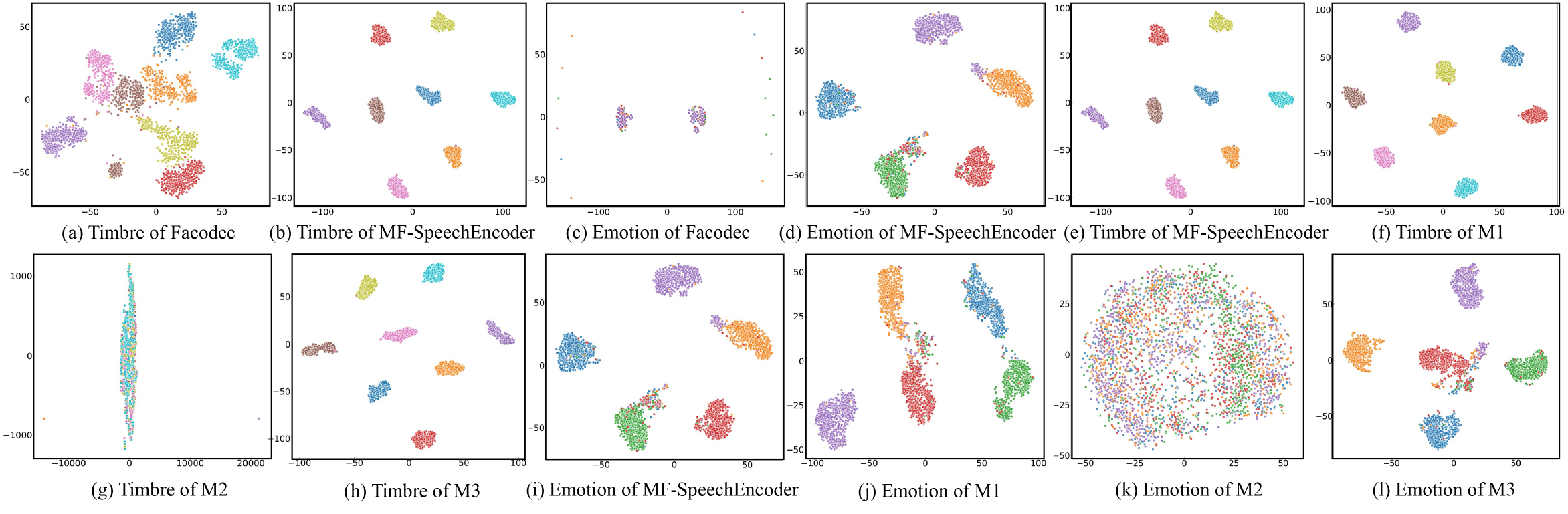}
  \caption{t-SNE visualization of timbre and emotion representation from four models.}
  \label{fig:tsne_compare_aba}
\end{figure*}

In order to further assess the high-purity and independence of the factors, we evaluated the results of MF-SpeechEncoder, as detailed in Table \ref{tab:DSFE_com}. In terms of mutual information, the performance is comparable to that of the baselines. However, in terms of the results of the target tasks (0.9979\%, 0.9296\%, 0.9593\%) and non-target tasks (0.2618\%, 0.0054\%, 0.1421\%), MF-SpeechEncoder has a more significant advantage. Although Facodec shows advantages in some metrics, MF-SpeechEncoder's t-SNE visualizations (Figure ~\ref{fig:tsne_compare_aba}.(a)-(d)) clearly demonstrate superior cluster separability and compactness, confirming robust disentanglement. Therefore, from multiple perspectives, MF-SpeechEncoder can produce more independent and pure factor representations. This effective separation of factors is foundational to MF-Speech's fine-grained and compositional control.

\begin{table}[h]
  \centering
  \begin{tabular}{lcccc}
    \hline
     Metrics & StyleVC & DDDM-VC & Facodec & \textbf{MF-SE}  \\ 
    \hline
    \midrule
     MI$_{te}\downarrow$  & 0.0070 & 0.0080  & \textbf{0.0063} & 0.0076     \\
     MI$_{tc}\downarrow$  & 0.0063 &  0.0079 & \textbf{0.0059} & 0.0061   \\
     MI$_{ec}\downarrow$  & 0.0080 & \textbf{0.0052} & 0.0197 & 0.0061      \\
    Acc$_t\uparrow$ &  0.9861  & 0.9882  &0.9939  & \textbf{0.9979}     \\
    Acc$_e\uparrow$ & 0.5168  & 0.2075 &0.2343 & \textbf{0.9296} \\
    Acc$_c\uparrow$   & 0.6661  & 0.8789  &0.0068   & \textbf{0.9593}    \\
    Acc$_{te}\downarrow$  & 0.5400  & 0.5579  &0.5932  &  \textbf{0.2618}      \\
    Acc$_{tc}\downarrow$  & 0.0321  & 0.0207 &0.0907 & \textbf{0.0054}    \\
    Acc$_{et}\downarrow$  & 0.7900  & \textbf{0.1339} &0.1596 & 0.2486 \\
    Acc$_{ec}\downarrow$ & 0.1111 & 0.0054 &\textbf{0.0021} & 0.0089 \\
    Acc$_{ct}\downarrow$ & 0.4268 & 0.9871 & 0.1704 & \textbf{0.1421}     \\
    Acc$_{ce}\downarrow$ & 0.4861 & 0.8118 & \textbf{0.2332} & 0.2529 \\
    \hline
  \end{tabular}
 \caption{The objective evaluation results of disentanglement. ``MF-SE" denotes the MF-SpeechEncoder.}
 \label{tab:DSFE_com}
\end{table}

\subsection {Ablation Study}

We conducted ablation studies on key components in the MF-SpeechEncoder and MF-SpeechGenerator modules separately to validate their respective contributions.

\textbf{MF-SpeechEncoder Component Analysis:} We ablated three components from the full MF-SpeechEncoder: Mutual Information constraints (M1: ${w/o_{MI}}$), contrastive learning (M2: ${w/o_{com}}$), and prosody priors (M3: ${w/o_{pro}}$).
\textbf{MI Constraints (${w/o_{MI}}$ - M1)(Figure ~\ref{fig:tsne_compare_aba}.(f), (j)):} After removing the MI constraint, the clustering boundaries of the emotion t-SNE become blurred, and the compactness of the clustering of timbre and emotion also decreases. This highlights the crucial role of MI in enhancing the discrimination and structural integrity of the factors. \textbf{Contrastive Learning (${w/o_{com}}$ - M2)(Figure ~\ref{fig:tsne_compare_aba}.(g), (k)):} The removal of contrastive learning  directly led to severe information entanglement. The t-SNE plots of timbre and emotion showed significant overlap and lacked clear boundaries when contrastive learning was removed, confirming that it plays a crucial role in achieving effective disentanglement. This observation suggests that minimizing mutual information between branches alone is insufficient; it is also essential to impose explicit constraints on each individual branch to preserve their respective factor-specific information. \textbf{Prosody Priors (${w/o_{pro}}$ - M3)(Figure ~\ref{fig:tsne_compare_aba}.(h), (l)):} Removing the prosody priors had a negative impact on the clustering of timbre and led to disordered emotion clusters. This highlights the crucial importance of the prosody priors for robust emotion modeling. As emotion representations become more accurately extracted, the mutual information constraints enable better disentanglement of timbre representations. Therefore, MF-SpeechEncoder demonstrates a high level of purity and stability in disentanglement, and its performance consistently outperforms all ablated variants.

\begin{figure}
  \centering
    \includegraphics[width=1.0\columnwidth]{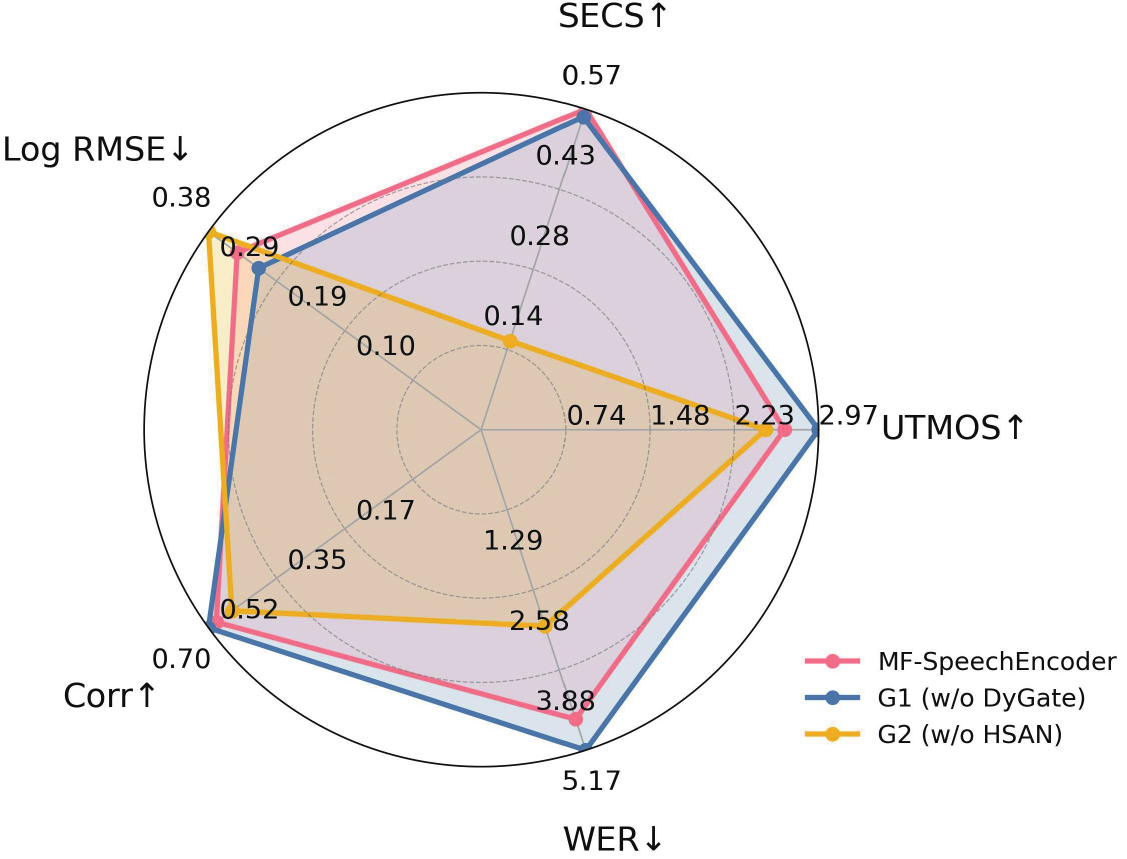}
    \caption{The ablation results of SpeechGenerator.}
  \label{fig:CSG_Abation}
\end{figure}

\textbf{SpeechGenerator Component Analysis:} We evaluated the impact of the dynamic fusion module (G1: $(w/o_{DyGate})$) and the HSAN (G2: $(w/o_{HSAN})$) on multi-factor compositional speech generation, with results in Figure \ref{fig:CSG_Abation}. \textbf{Dynamic Fusion ($(w/o_{DyGate})$ - G1):} Removing dynamic fusion slightly reduced timbre similarity (SECS from 0.5685 to 0.5551) and increased the word error rate (WER from 4.76\% to 5.17\%). Although it showed marginal improvements in UTMOS, Log RMSE, and Corr, the differences were minimal. Given that WER and SECS are more critical for evaluating controllability, these results suggest that dynamic fusion plays a positive role in preserving the consistency of both timbre and content. \textbf{HSAN ($(w/o_{HSAN})$ - G2):} Removing HSAN had a more pronounced negative impact. Except for WER, all other metrics deteriorated significantly (SECS dropped to 0.1576, Corr dropped to 0.64, and Log RMSE increased to 0.38). The relatively better WER can be attributed to the fact that, without HSAN, the fused features are used directly for generation, eliminating the need for additional control injections of timbre and emotion. This allows the model to focus more on content representation, thereby improving WER. However, our primary focus is on fine-grained and compositional control. Despite the improved WER, the ability to control timbre and emotion becomes significantly weaker without HSAN. Therefore, HSAN remains an indispensable component, playing a crucial role in ensuring style consistency and controllable synthesis. 
Overall, dynamic fusion and HSAN play complementary and crucial roles: Dynamic fusion helps achieve multi-factor coordination and content integrity, while HSAN is the key to realizing powerful style expression and control.

\section{Conclusion}
In this paper, we introduce MF-Speech, a framework addressing factor entanglement and coarse control in controllable speech generation. Our solution has two core components: MF-SpeechEncoder, acting as a factor purifier, uses a multi-objective strategy to produce highly disentangled discrete representations of content, timbre, and emotion. MF-SpeechGenerator, as a speech conductor, leverages these purified factors via a dynamic fusion module and HSAN to enable fine-grained, compositional control. Extensive experiments confirm our approach's effectiveness, showing MF-Speech significantly outperforms existing methods in content fidelity and style similarity. Future work will further improve synthesis quality and generalization across diverse speaker and emotion conditions.

\appendix

\section{Acknowledgments}
This work was supported in part by the National Natural Science Foundation of China under Grants 
62121002,62372423,62072421,62476013 and was also supported by the Fundamental Research Funds for the Central Universities WK2100250070

\bibliography{aaai2026}

\end{document}